\documentclass{elsart}

\usepackage[hiresbb,final]{graphicx}
\usepackage{amsmath}
\newcommand{\bfd}{\mathbf{d}}
\newcommand{\bff}{\mathbf{f}}
\newcommand{\bfF}{\mathbf{F}}

\newcommand{\bfN}{\mathbf{N}}
\newcommand{\bfT}{\mathbf{T}}
\newcommand{\bfr}{\mathbf{r}}

\newcommand{\bfv}{\mathbf{v}}
\newcommand{\bfx}{\mathbf{x}}
\newcommand{\bfy}{\mathbf{y}}
\newcommand{\bfX}{\mathbf{X}}
\newcommand{\bfzero}{\mathbf{0}}
\newcommand{\bfxi}{{\boldsymbol{\xi}}}
\newcommand{\calK}{{\mathcal{K}}}
\newcommand{\calT}{{\mathcal{T}}}

\newcommand{\bfnabla}{\pmb\nabla}
\newcommand{\divergence}{\operatorname{div}}
\newcommand{\laplace}{\Delta}
\newcommand{\scalarmult}{\!\boldsymbol{\cdot}\!}

\newcommand{\pascal}{\nobreak\mbox{Pa}}
\newcommand{\kelvin}{\nobreak\mbox{K}}

\newcommand{\meter}{\nobreak\mbox{m}}

\newcommand{\viscosity}{\eta}

\newcommand{\stress}{\sigma}

\newcommand{\Peclet}{\textit{Pe}}
\newcommand{\order}[1]{\mathcal{O}(#1)}
\newcommand\scriptbox[2]{{\makebox[#1]{\hss$\scriptstyle #2$\hss}}}

\begin{document}
\begin{frontmatter}
\title{Accumulating Particles at the Boundaries of a Laminar Flow}

\author[A,P]{Michael Schindler},
\author[A]{Peter Talkner},
\author[A]{Marcin Kostur},
\author[A]{Peter H\"anggi\corauthref{cor}}
\corauth[cor]{\texttt{Peter.Hanggi@Physik.Uni-Augsburg.DE}\ (Peter H\"anggi)}
\address[A]{Institut f\"ur Physik, Universit\"at~Augsburg,
            Universit\"atsstra{\ss}e~1, D--86135~Augsburg, Germany}
\address[P]{Laboratoire~PCT, UMR~``Gulliver''
            CNRS-ESPCI~7083,\\ 10~rue~Vauquelin, 75231~Paris cedex~05}


\begin{abstract}
The accumulation of small particles is analyzed in stationary flows through
channels of variable width at small Reynolds number. The combined influence of
pressure, viscous drag and thermal fluctuations is described by means of a
Fokker-Planck equation for the particle density. It is shown that for extended
spherical particles the shape of the fluid domain gives rise to inhomogeneous
particle densities, thereby leading to particle accumulation and corresponding
depletion. For extended spherical particles, conditions are specified that lead
to inhomogeneous densities and consequently to regions with particle
accumulation and depletion.

\end{abstract}

\end{frontmatter}


\section{Introduction}
\label{sec:intro}%
During recent years the investigation of microfluidic systems has increasingly
attracted attention, being boosted by its promising and powerful so-called
``lab-on-a-chip''~\cite{loac1,loac2,loac3} applications. A standard task that
such a device should be able to perform is the separation of small objects
immersed in a fluid according to specific properties of these objects like size
or form. For this purpose, several mechanisms have been suggested. The proposed
methods are either based on downsized conventional laboratory tools like tubes,
pumps and valves, or they employ effects that only function in the microfluidic
regime. In combination with particular geometric
constraints~\cite{gc1,gc2,gc3,gc4,gc5} and various ways of external forcing such
as by electric fields~\cite{ef1,ef2,ef3,condmat}, thermal gradients
\cite{therm1,therm2} or acoustic streaming induced by strong ultrasound waves
\cite{as1,as2,as3}, thermal fluctuations play a key-role in many of these
separation techniques \cite{gc1,gc2,gc3,gc4,chiral,KetReiHanMul00}.

Of great practical and principle interest are sorting methods that exclusively
utilize inhomogeneities  of flows in constrained geometries in combination with
the omnipresent thermal noise. In particular, there are no external forces
acting on the particle with these methods.  The drift ratchet provides a typical
example of this kind of device~\cite{KetReiHanMul00,halle1,halle2,halle3}. It
consists of a large number of identical channels in a membrane, connecting two
reservoirs both filled with water. The radius of each channel varies
periodically in an asymmetric way. A carrier fluid, in most cases water, is
periodically pumped to and fro. Although no water is transported on average,
particles immersed in the water in general move to one side with a non-vanishing
average velocity. Separation becomes possible because the particle velocity and
even its direction depend on the size of the particle
\cite{KetReiHanMul00,halle1}. As in conventional
ratchets~\cite{ratchets1,ratchets2,ratchets3,ratchets4,ratchets5,ratchets6} the
combined action of the periodic but asymmetric pore shape, breaking the
left--right symmetry, the thermal fluctuations and the periodic pumping which
drives the system out of thermal equilibrium are necessary ingredients for the
observed effect~\cite{KetReiHanMul00}.

\begin{figure}[tb]
  \centering%
  \includegraphics{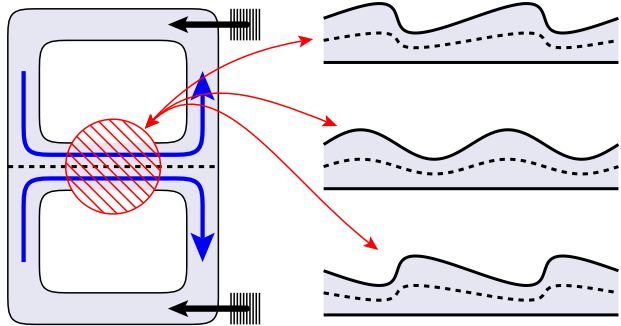}%
  \caption{(Color online) Sketch of the channel geometry. The middle part
  contains an asymmetric channel, which is doubly closed by two loops. The
  dashed line indicates the streamline connecting the stagnation points of the
  flow and separating the two flow chambers. The geometry may represent either
  closed channels or a wetting pattern on a flat substrate. The flow driving is
  exemplified to be acoustically driven as indicated by the two blocks of
  parallel lines representing surface-acoustic waves which propagate in the
  directions of the arrows~\cite{eightshape,Frommelt07}. }%
  \label{f0}
\end{figure}

The type of device studied in the present paper differs in several aspects from
a drift ratchet. It does not require an oscillating flow, which is difficult to
generate and maintain experimentally~\cite{halle1,halle2,halle3}, but operates
with a stationary flow field. The flow is confined to a channel with variable
width -- as for a drift ratchet -- but with ends connected to each other in some
way, leading e.g. to a ring-like or ``figure eight'' geometry, cf.\
Fig~\ref{f0}. Spherical particles will then accumulate in particular regions of
the channel. In a previous experimental work the figure eight geometry was
realized to separate particles immersed in a fluid by means of electric
fields~\cite{eightshape,Frommelt07}. In this experiment, a steady flow of the
fluid was maintained by surface acoustic waves leading to acoustic
streaming~\cite{as1}.

For the sake of simplicity we restrict our analysis to two spatial dimensions.
The relevant mechanisms put forward in this work leading to an accumulation do
not depend on the two-dimensional character and will work as well in three
dimensions. In the experimental realization~\cite{eightshape,Frommelt07} the
flow channels are bounded by a rigid substrate only at the bottom, the remaining
boundary is a free interface between water and air. A reduction of this very
situation to two dimensions is not obvious. To still include the effects of
different types of boundaries in the two-dimensional model we consider two
cases, one with two no-slip boundaries corresponding to two impenetrable walls
confining the fluid and the other one with a no-slip and a perfect slip
boundary. The latter boundary mimics a free boundary with dominant surface
tension such that the flow only leads to a negligible deformation of the
equilibrium shape.%

The paper is organized as follows. In Section~\ref{sec:flow} we introduce the
flow field which is then used for transporting particles. In
Section~\ref{sec:densities} we analyze the hydrodynamic and random forces on the
particles and formulate the long-time particle accumulation pattern in terms of
a stationary Fokker--Planck equation. Section~\ref{sec:effects} makes a
distinction between possible accumulations caused by a volume and a boundary
effect. In pressure-driven channel flows only the latter may occur, which is
then investigated numerically in Section~\ref{sec:numerics}.

\section{The flow fields}
\label{sec:flow}

We consider stationary incompressible flows in the limit of vanishing Reynolds
numbers. The velocity~$\bfv(\bfx)$ and pressure~$p(\bfx)$ fields are solutions
of the stationary Stokes equations,
\begin{gather}
  \label{eq:stokes}
  \bfzero = - \bfnabla p(\bfx) + \viscosity\laplace\bfv(\bfx) + \bff(\bfx), \\
  \label{eq:incompress}
  0 = \divergence\bfv(\bfx),
\end{gather}
where $\viscosity$~denotes the viscosity of the fluid and $\bff$~is a
conservative externally applied body force which gives rise to the flow fields
$p(\bfx)$ and $\bfv(\bfx)$. The velocity field is subject to the \emph{kinematic
boundary condition}, i.\,e.\ for immobile boundaries,
\begin{equation}
  \label{eq:kinematic}
  \bfv\scalarmult\bfN = 0,
\end{equation}
representing the fact that all boundaries of the fluid domain~$\partial\Omega$
are material lines of the flow. $\bfN$~denotes the normal vector. At immobile
sticky walls, also the velocity components parallel to the tangent
vectors~$\bfT^{(\alpha)}$ vanish,
\begin{equation}
  \label{eq:noslip}
  \bfv\scalarmult\bfT^{(\alpha)} = 0\quad\text{for all $\alpha$.}
\end{equation}
For a free boundary, such as the interface between water and air, the tangential
velocity at the boundary is determined by the mechanical stress balance. For the
velocity field, this reduces to the \emph{perfect slip condition}
\begin{equation}
  \label{eq:slip}
  T_i^{(\alpha)} \left(\frac{\partial v_i}{\partial x_j} + \frac{\partial v_j}{\partial
  x_i}\right) N_j = 0\quad\text{for all $\alpha$.}
\end{equation}
In general, at a free boundary the shape of the interface between two fluids
depends itself on, and reacts back to, the flow and pressure fields in the
fluid. Here we restrict ourselves to flows in a given domain on the boundaries
of which the kinematic boundary condition (\ref{eq:kinematic}) and either the
no-slip (\ref{eq:noslip}) or the perfect slip condition (\ref{eq:slip}) hold.
The full description of free fluid surfaces with surface tension can be found in
references~\cite{SchTalHanPoF,Schindler06}. In the numerical examples in
Sec.~\ref{sec:numerics} below, we will find that the type of the boundary
condition has strong influence on the particle accumulation taking place near
the boundary.

\section{Hydrodynamic particle transport}
\label{sec:densities}

The flow fields solving the Stokes equations~\eqref{eq:stokes} and
\eqref{eq:incompress} in combination with the boundary conditions
\eqref{eq:kinematic} and \eqref{eq:noslip} or \eqref{eq:slip} are used to advect
small spherical particles. If such a particle is a point-particle at
position~$\bfX(t)$, it cannot be distinguished from the fluid material at this
point. It is therefore transported with the velocity of the fluid itself,
$\dot\bfX(t) = \bfv(\bfX(t))$.

The motion of an extended spherical particle with small but non-vanishing
radius~$R$ about~$\bfX(t)$, however, is qualitatively different from that of a
point-particle. The force on such a particle is the integral of the fluidic
stress over the particle's oriented surface, denoted by $\bfF_\bfv(t)$ and
$\bfF_p(t)$ for the viscous and the pressure contributions of the stress,
respectively. An additional random force~$\bfF_\bfxi(t)$ takes thermal
fluctuations into account. For a small particle of mass~$m$, the inertial force
is negligible~\cite{Purcell77}, resulting in an overdamped motion, described by
the Langevin equation
\begin{equation}
  \label{eq:langevin_overdamped}
  0 = m\ddot\bfX(t) = \bfF_\bfv(t) + \bfF_p(t) + \bfF_\bfxi(t).
\end{equation}

The pressure contribution~$\bfF_p(t)$ is obtained by expanding the pressure
field in a Taylor series around the particle center~$\bfX$ and integrating
over the particle surface~$S_R(\bfX)$,
\begin{equation}
  \label{eq:pressureforce}
  \bfF_p
  = -\oint\limits_{\makebox[1em]{\hss$\scriptstyle S_R(\bfX)$\hss}} p \bfN\,dA
  = -\frac{4\pi}{3} R^3 \bfnabla p(\bfX)
    -\order{R^5}.
\end{equation}

The viscous contribution~$\bfF_\bfv(t)$ cannot be readily obtained, since the
particle alters the velocity field: It poses a spherical no-slip boundary to the
surrounding fluid. In the Stokes regime, in which the acceleration field of the
fluid caused by the movement of the particle can be neglected, we may employ
Fax\'en's theorem of translational motion to describe this effect
\cite{Pozrikidis92,Schindler06}. The force~$\bfF_\bfv(t)$ is then given in terms
of a modified velocity field~$\tilde\bfv$, which is evaluated at the center of
the particle,
\begin{equation}
  \label{eq:faxen}
  \bfF_\bfv
  = 6\pi\viscosity R \bigl[- \dot\bfX + \tilde\bfv(\bfX)
    + \laplace\tilde\bfv(\bfX)\,R^2/6 \bigr].
\end{equation}
In contrast to the true velocity field $\bfv$ which describes the flow in the
presence of the particle, the auxiliary field $\tilde{\bfv}$ is also defined
within the region which is covered by the particle. Still, it contains all
perturbations caused by the particle, cf.\ Appendix~\ref{appendix} for the
definition and Ref.~\cite{Schindler06} for a detailed discussion. Unfortunately,
determining~$\tilde\bfv$ is as elaborate as solving the full stationary problem
in the presence of the particle. In the next section we can, however, make use
of the fact that $\tilde\bfv$~is a solenoidal vector field.

The third contribution to the force on a particle in
Eq.~\eqref{eq:langevin_overdamped}, i.e.\ the fluctuating force~$\bfF_\bfxi$,
reads at thermodynamic equilibrium~\cite{Einstein1,Einstein2,Einstein3},
\begin{equation}
  \label{eq:velocdiff}
  \bfF_\bfxi(t) = \sqrt{2d\,}\, \bfxi(t),
\end{equation}
with the noise strength $d = 6\pi\viscosity R\, k_\text{B} T$, and
$\bfxi(t)$~being Gaussian white noise,
\begin{gather}
  \bigl\langle\xi_i(t)\bigr\rangle = 0,\\
  \bigl\langle\xi_i(t)\xi_j(s)\bigr\rangle = \delta_{ij}\,\delta(t-s).
\end{gather}
We neglect deviations from the equilibrium expression of the
fluctuation--dissipation theorem, which become relevant for large velocity
gradients only, see Refs.~\cite{Rubi1,Rubi2}, and references therein. Further,
we assume a dilute solution of particles such that hydrodynamic
particle--particle interactions can be ignored in the Langevin equation
(\ref{eq:langevin_overdamped}). The overdamped Langevin
equation~\eqref{eq:langevin_overdamped}, solved for the particle velocity is
then equivalent to a Fokker--Planck equation for the probability
density~$\rho(\bfX, t)$ of the particle centers,
\begin{equation}
  \partial_t\rho(\bfx,t)
  = -\divergence\bigl(\rho(\bfx)\,\bfd(\bfx)\bigr)
  + D \laplace \rho(\bfx).
\end{equation}
The diffusion constant is given by the Sutherland--Einstein relation
\begin{equation}
  D = \frac{d}{(6\pi\viscosity R)^2} = \frac{k_\text{B}T}{6\pi\viscosity R},
\end{equation}
and the drift velocity $\bfd$~reads, combining \eqref{eq:pressureforce} and
\eqref{eq:faxen},
\begin{equation}
  \label{eq:drift}
  \bfd(\bfx) =
       \tilde\bfv(\bfx) + \frac{R^2}{6}
       \laplace\tilde\bfv(\bfx)
      -\frac{2R^2}{9\viscosity} \bfnabla p(\bfx)
      +\order{R^5}.
\end{equation}
The remaining term~$\order{R^5}$ comes from the Taylor expansion in
Eq.~\eqref{eq:pressureforce}. The viscous terms containing the modified velocity
field~$\tilde\bfv$ are exact. In the present paper, we focus on the long-time
limit of the particle distribution, which is governed by the condition
$\partial_t\rho(\bfx,t) = 0$, leading to the stationary Fokker--Planck equation
\begin{equation}
  \label{eq:fp}
  0
  = -\divergence\bigl(\rho(\bfx)\,\bfd(\bfx)\bigr)
  + D \laplace \rho(\bfx).
\end{equation}
%


\section{Volume- versus boundary-accumulation}
\label{sec:effects}%
As an \emph{accumulation} of particles we refer to an inhomogeneous distribution
of particles~$\rho(\bfx)$. First, we note that the uniform distribution
$\rho(\bfX) = \textit{const}$ is a possible solution of the stationary
Fokker--Planck equation~\eqref{eq:fp} in case of a solenoidal drift field,
$\divergence(\bfd)=0$ \cite{condmat,chiral}. Whether this solution is the
physically realized one, depends only on the boundary conditions. Deviations
from a uniform stationary particle density are caused either by particular
boundary conditions which impose a non-vanishing gradient of the particle
density at the boundary, or a non-vanishing divergence of the vector
field~$\bfd(\bfx)$. We refer to the former and the latter case as \emph{boundary
effect} and \emph{volume effect}, respectively.

\subsection{Volume effect}%
\label{sec:volume}%

The volume accumulation mechanism takes place for a drift velocity having a
non-vanishing divergence, $\divergence(\bfd)\neq0$. The homogeneous distribution
$\rho(\bfX) = \textit{const}$ then is not a solution of equation~\eqref{eq:fp},
irrespectively of any boundary condition. Whether the drift field is solenoidal,
depends on the shape of the particle. A particle of complicated shape generally
gives rise to a drift field with non-vanishing divergence. Examples can be found
in the literature
\cite{volumeeffect1,volumeeffect2,volumeeffect3,volumeeffect4,volumeeffect5,chiral}.

In the special case of a spherical particle, however, the drift velocity
$\bfd(\bfX)$~is given by equation~\eqref{eq:drift}. Although the velocity
field~$\tilde\bfv(\bfX)$ is not known explicitly, it is known to be
divergence-free. The same is true for its Laplacian. Only the pressure
gradient may lead to a non-vanishing divergence of~$\bfd(\bfx)$. Taking the
divergence of Eq.~\eqref{eq:drift} one obtains in leading order the Laplacian of
the pressure~$p$, which can be determined by the Stokes
equation~\eqref{eq:stokes} to yield
\begin{equation}
  \divergence(\bfd)
  = \frac{2R^2}{9\viscosity} \laplace p
  = \frac{2R^2}{9\viscosity} \divergence(\bff).
\end{equation}
In a long modulated channel, where the flow is driven either by a pressure
difference, by a constant force, or by an inflow velocity profile, the drift
velocity~$\bfd(\bfx)$ is indeed solenoidal. For these situations we state that
all volume effects vanish exactly, and inhomogeneities in the particle density
can only be caused by the boundaries.

\subsection{Boundary effect}
\label{sec:boundary}

The boundary condition for the particle density in the fluid is given by the
condition of vanishing normal particle flux,
\begin{equation}
  \label{eq:no_flux}
  0 = \bfN\scalarmult(\rho\,\bfd - D\,\bfnabla\rho),
\end{equation}
which guarantees that particles cannot cross the boundary. In case of a
vanishing normal drift velocity~$\bfN\scalarmult\bfd = 0$ only the uniform
distribution $\rho(\bfx) = \textit{const}$ is allowed. Vice versa, an
inhomogeneous particle distribution can only emerge, if the normal drift at the
boundary does not to vanish.

\begin{figure}[t]%
  \centering
  \includegraphics{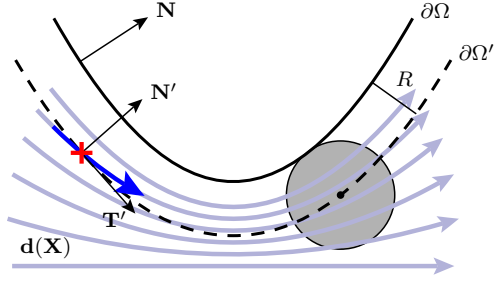}%
  \caption{(Color online) The boundaries of the fluid domain~$\Omega$ and of the
  accessible region~$\Omega^\prime$ for the particle center. Their distance
  equals the particle radius~$R$ everywhere. At the boundary~$\partial
  \Omega^\prime$ the normal component of the drift velocity, and thus also of
  the drag force, are not zero. This is exemplified at the point marked by a
  cross.}%
  \label{fig:bdstreamlines}%
\end{figure}

The velocity field $\tilde\bfv$, which is the leading-order term in the drift
velocity~\eqref{eq:drift}, satisfies the kinematic boundary
condition~\eqref{eq:kinematic} at the boundary of the fluid domain. As it does
not provide a normal component, one might expect that neither a volume effect
nor a boundary accumulation is achieved from the leading-order term in
Eq.~\eqref{eq:drift}. A close look on a spherical particle in the vicinity of a
boundary, however, reveals that there is indeed such a normal component.
Relevant in Fax\'en's theorem~\eqref{eq:faxen} is the modified velocity
field~$\tilde\bfv$ at the center of the particle, which cannot touch the
physical boundary but stays at least one~radius apart from it. The no-flux
boundary condition~\eqref{eq:no_flux} must therefore be imposed on an effective
boundary~$\partial\Omega^\prime$ which has constant distance~$R$ from the
physical boundary~$\partial\Omega$ of the fluid. Both are sketched in
Fig.~\ref{fig:bdstreamlines}. Neither the streamlines of the flow nor the
field-lines of the particle drift velocity need to have a constant distance from
the boundary. Instead, some of the field-lines cross the
boundary~$\partial\Omega^\prime$ of the region which is accessible to the
particle centers. In Fig.~\ref{fig:bdstreamlines} one of these crossing points
is marked. There, the normal component of~$\bfd(\bfX)$ is positive, i.\,e.\
pointing to the boundary whence leading to a deposition of particles at the
boundary. Accordingly, a negative normal component results in a depletion zone.
Both effects will be found in the numerical simulations below, see
Fig.~\ref{fig:density}.%

The given argument in the foregoing paragraph is a geometrical explanation for
the leading-order terms of the hydrodynamic interaction between particle and
wall. A general theory of hydrodynamic interactions in asymmetric geometries is
presently not available. A known effect of the hydrodynamic interaction is a
decrease of the fluid velocity induced by the presence of the particle.
Additionally, one obtains the Saffman lift force~\cite{Saff} which drives the
particle away from the boundary. The magnitude of this force is proportional to
the squared particle radius and to the mismatch between the true particle
velocity and the unperturbed fluid velocity. Reference~\cite{LonKleBuc04}
provides explicit expressions for the correction terms from hydrodynamic
interaction between particle and boundary. The force densities given there
comprise a component, diverging at the effective
boundary~$\partial\Omega^\prime$, corresponding to the hard-core interaction.
The next leading-order term in the particle radius $R$ is the random force,
being proportional to~$R^{1/2}$. The tangential force leading to the slowing
down of the particle is proportional to~$R$, and the Saffman lift force scale
as~$R^2$.

About the normal projection of the second-order terms in the drift
velocity~$\bfd(\bfx)$, being proportional to $R^2$ in Eq.~\eqref{eq:drift}, we
cannot say much. Generally, they may give rise to a boundary effect. The
magnitude of this effect, however, scales quadratically with the particle
radius~$R$ and is therefore expected to be smaller than the hard-sphere boundary
effect. Furthermore, for pressure-driven flows in long channels, the pressure
gradient at the boundary is oriented along the channel rather than normal to it.

\section{Numerical examples in modulated channels}
\label{sec:numerics}%
The aim of this paper is to provide a qualitative picture of the influence of
the domain shape on the particle density. In order to achieve this goal we have
to restrict the following numerical analysis to the leading-order terms of the
forces. These are the geometrical interaction between particles and wall, as
depicted in Fig.~\ref{fig:bdstreamlines}, and the random forces. By omitting
terms which are quadratic in the particle radius, we may also replace the
unknown velocity field~$\tilde\bfv$ by the unperturbed velocity~$\bfv$.

Using these simplifications, the full interaction between particle and wall is
replaced by a hard-core interaction preventing the particle from penetrating the
wall. We are aware that this approximation may be quite crude  for the
quantitative description of a real particle in a real fluid. However, the
analysis above shows that the only terms which lead to an accumulation are the
normal forces on the particles in the vicinity of the boundary. In the hard-core
model, this happens instantaneously as the particle touches the wall. The force
is localized on the effective boundary~$\partial\Omega^\prime$. Corrections to
the force, as detailed in Ref.~\cite{LonKleBuc04} smear out the force such that
the particle is slowed down already in the vicinity of the wall. This will not
alter the qualitative picture obtained here by the hard-core interaction.
Whether this approximation is really justifiable, can be decided only by
extensive numerical calculations of the hydrodynamic interactions, a task which
is beyond the scope of the present study.

\begin{figure}%
  \centering
  \medskipamount=1ex
  \includegraphics{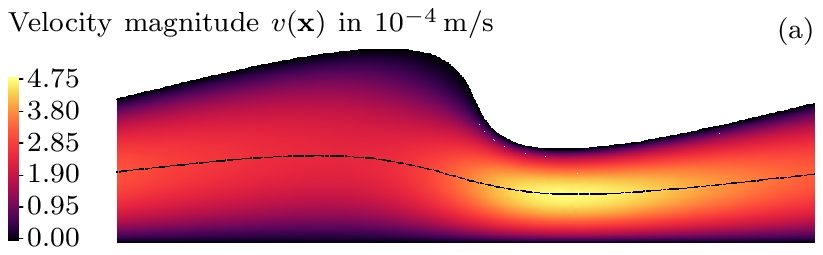}\medskip\par
  \includegraphics{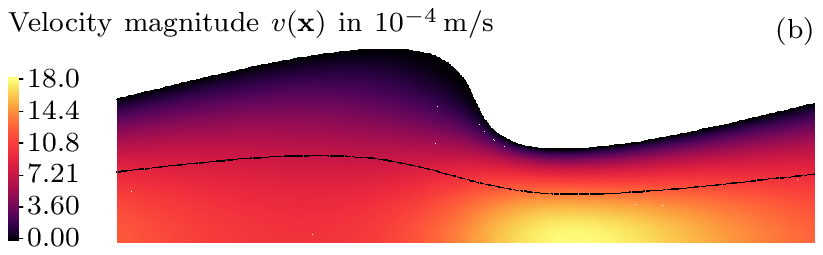}%
  \caption{(Color online) Example-flows in periodically continued
  two-dimensional channels, modulated according to Eq.~\eqref{eq:asymmetry} with
  asymmetry parameter $a=0.8$. In panel~(a) both boundary conditions are
  no-slip, while in (b)~the straight boundary carries a perfect-slip condition.
  The flow direction is from left to right, driven by a pressure difference of
  $\Delta p = 1\,\pascal$ per length $L=10^{-4}\meter$ of the unit cell. The
  viscosity is that of water. The black lines represent streamlines.}%
  \label{fig:flow}%
\end{figure}

Next, we solve the Fokker--Planck equation
\begin{equation}
  \label{eq:num_fp}
  0 = - \bfv\scalarmult\bfnabla\rho + D\laplace\rho
  \quad\text{in $\Omega^\prime$,}
\end{equation}
with the boundary condition
\begin{equation}
  \label{eq:num_fp_bd}
  0 = \bfN^\prime \scalarmult (\rho\,\bfv - D\bfnabla\rho)
  \quad\text{on $\partial\Omega^\prime$.}
\end{equation}
The vector $\bfN^\prime$ denotes the normal vector of the effective
boundary~$\partial\Omega^\prime$. The velocity field~$\bfv$ is taken from the
numerical solution of the Stokes equations in~$\Omega$ with the boundary
conditions from Sec.~\ref{sec:flow}, the domain~$\Omega$ being a periodic
two-dimensional channel of modulated width. Its two boundaries have different
geometries, one is straight, while the other is curved, see Fig.~\ref{fig:flow}.
We choose the shape of the curved boundary to be given by the periodic
function~$g$, which is implicitly defined by the relation
\begin{equation}
  \label{eq:asymmetry}
  g(z) = \sin\bigl(2\pi z - a\,g(z)\bigr).
\end{equation}
%
\begin{figure}%
  \centering
  \includegraphics{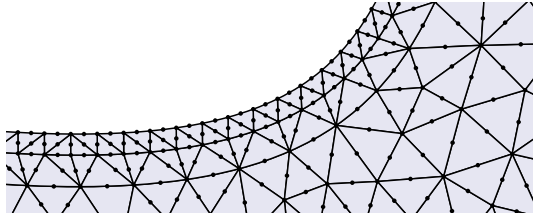}%
  \caption{(Color online) A detail of the computational mesh near the boundary. The outermost
  layer of finite elements represents the zone which cannot be entered by the
  particle centers. The element sides of this layer constitute the two
  boundaries $\partial\Omega$~and $\partial\Omega^\prime$, compare with
  Fig.~\ref{fig:bdstreamlines}.}%
  \label{fig:boundarymesh}%
\end{figure}%
\begin{figure}%
  \centering
  \medskipamount=1ex
  \includegraphics{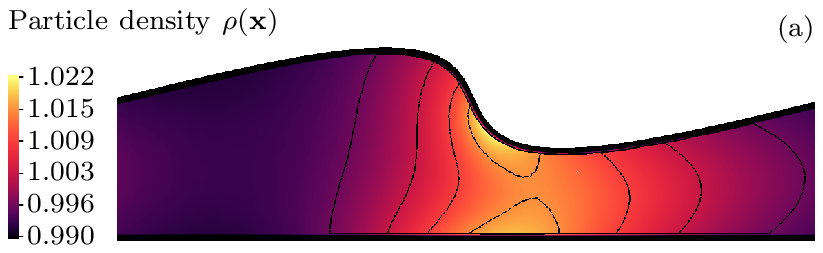}\medskip\par
  \includegraphics{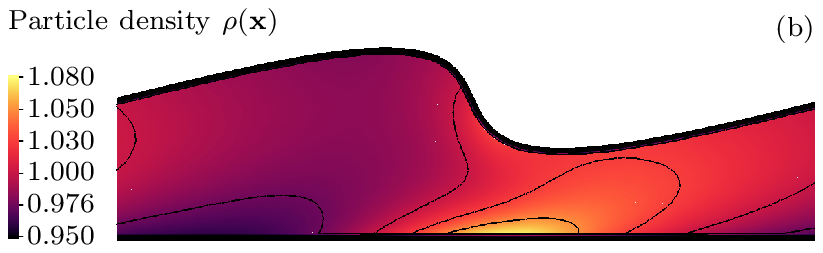}%
  \caption{(Color online) Example-densities of particle centers in the flows of
  Fig.~\ref{fig:flow}, normalized to an average value of unity. The densities
  have been calculated as the solution of the Fokker--Planck
  equation~\eqref{eq:fp} with $\Delta p/T=3.3\times10^{-6}\,\pascal/\kelvin$
  which corresponds to a P\'eclet number of $\Peclet\approx 16$. At the
  boundaries, the black zone with zero density indicates the area which the
  particle centers cannot enter. It is one particle radius wide, here
  $R=10^{-2}L$. The thin black lines indicate level lines of the density.}%
  \label{fig:density}%
\end{figure}%
The number $a$~parametrizes the asymmetry of the shape, yielding a sinusoidal
shape for $a=0$, while shapes with $a\neq 0$ have alternating steep and flat
flanks. For $a>0$ the flat flanks have positive and the steep flanks have
negative inclination, and vice versa for~$a<0$. The values $g(z)$ can be
evaluated iteratively, starting with $\sin(2\pi z)$.

The parameter regime of equation~\eqref{eq:num_fp} is characterized by the
dimensionless \emph{P\'eclet number}, which expresses the ratio of advective to
diffusive transport. We here define the global P\'eclet number as
\begin{equation}
  \Peclet
  = \frac{\bar{x}\bar{v}}{D}
  = \frac{3\pi L^2}{128\,k_\text{B}}\:R\:\frac{\Delta p}{T}.
\end{equation}
For the stationary particle distribution, there is no difference between a
weakly driven system and one at a high temperature, as long as the ratio $\Delta
p/T$ is the same. In the numerical calculations for several different
temperatures, we therefore used the same numerically obtained velocity field,
driven by a unit pressure difference $\Delta p = 1\,\pascal$ along the channel
length $L=10^{-4}\,\meter$.

The computational mesh for the numerical solutions has to provide both
boundaries, $\partial\Omega$~and $\partial\Omega^\prime$.
Figure~\ref{fig:boundarymesh} shows a detail of the employed mesh near a
boundary. The outermost layer of finite elements build the zone of constant
width~$R$ which cannot be entered by the particle centers. This zone can be
identified in Fig.~\ref{fig:density} as the zone near the boundary, showing a
vanishing particle density.

\subsection{Longitudinal accumulation}

Figure~\ref{fig:flow} presents two velocity fields in the unit cell of a channel
with asymmetry parameter $a=0.8$. They differ by the boundary condition at the
straight wall: Panel (a) presents no-slip and (b) perfect slip boundary
conditions. The corresponding solutions of the stationary Fokker--Planck
equation~\eqref{eq:num_fp} with boundary condition~\eqref{eq:num_fp_bd} are
illustrated in Fig.~\ref{fig:density}. The qualitative form of the stationary
distributions is quite intuitive. Left of the bottleneck we find an
accumulation zone of the particles. Resulting from the main flow direction
from left to right, the particles are concentrated at the left side, just
upstream in front of the bottleneck. This result can be understood by means of
Fig.~\ref{fig:bdstreamlines}. Left and right of the bottleneck we find regions
where streamlines cross the effective boundary~$\partial\Omega^\prime$, leading
to a non-vanishing drift component in normal direction. This normal component
results in the accumulation that we see in Fig.~\ref{fig:density} for
$\Peclet\approx 16$. This corresponds to a slightly advection-dominated
transport of particles. For smaller P\'eclet numbers, the diffusion becomes more
dominant, resulting in a smoother particle distribution than in
Fig.~\ref{fig:density}. For larger P\'eclet numbers, the maxima of the
distribution near the boundaries become more pronounced.

\subsection{Influence of the boundary conditions}

Comparing the particle densities in Fig.~\ref{fig:density} for different
boundary conditions of the velocity field, one finds the extremal values to be
more pronounced in the case with a perfect-slip condition. The reason for this
difference is that the slip condition allows a larger velocity at the boundary,
giving rise also to a larger normal component, which then causes the
accumulation effect. We expect that in general free surfaces, providing perfect
slip boundary conditions, cause larger accumulation effects than sticky walls,
as long as the particles do not leave the fluid nor cause major deformations of
the free surface.

\subsection{Perpendicular accumulation and sorting}

The two accumulation patterns in Fig.~\ref{fig:density} do not only differ with
respect to their magnitude. So far, we have only looked at the inhomogeneity
\emph{along} the channel orientation, by stating that there are concentration
and depletion zones in front of the bottleneck and behind, respectively. For a
small channel as in experimental applications also the distance of the
accumulation centers will be prohibitively small. A performance characteristic
of much more interest is the accumulation \emph{perpendicular} to the channel
orientation. This may be achieved by a branching channel where the flow is split
along the streamline that ends in a stagnation point. Such a structure has been
experimentally realized as a eight-shaped geometry and used for separating
particles with external electric forces \cite{eightshape}. Particles immersed in
this flow do not generally follow the streamlines and thus can cross the
\emph{separating streamline} and pass from one basin into the other. This leads
to a \emph{relative accumulation} of particles in one of the basins, if they
were originally equally distributed.

The quality of separation is characterized by the amounts of particles that are
above and below the separating streamline indicated in Fig.~\ref{fig:flow}. For
this purpose we separately integrate the particle density in the two regions
$\Omega^\prime_{+}$~and $\Omega^\prime_{-}$, representing the accessible regions
above ($+$) and below~($-$) the separating streamline, respectively. The
resulting two probabilities per area $P_{+}/\bigl|\Omega^\prime_{+}\bigr|$ and
$P_{-}/\bigl|\Omega^\prime_{-}\bigr|$ of finding a particle above or below the
streamline can be used to provide a measure for the relative accumulation of
particles in one of the basins. We use the quantity
\begin{equation}
  \label{eq:ratio}
  r :=
  \frac{P_{+}}{\bigl|\Omega^\prime_{+}\bigr|}
  \biggm/
  \frac{P_{-}}{\bigl|\Omega^\prime_{-}\bigr|}
\end{equation}
as a measure for the relative accumulation in the upper basin, which is the one
with the curved boundary. The denominators $\bigl|\Omega^\prime_{+}\bigr|$~and
$\bigl|\Omega^\prime_{-}\bigr|$ represent the volumes of the regions above and
below the separating streamline, respectively.%
\begin{figure}%
  \centering
  \includegraphics{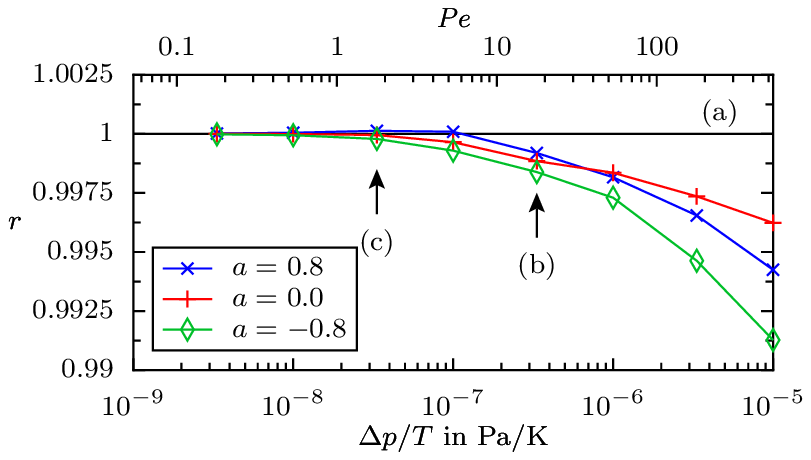}\medskip\par
  \includegraphics{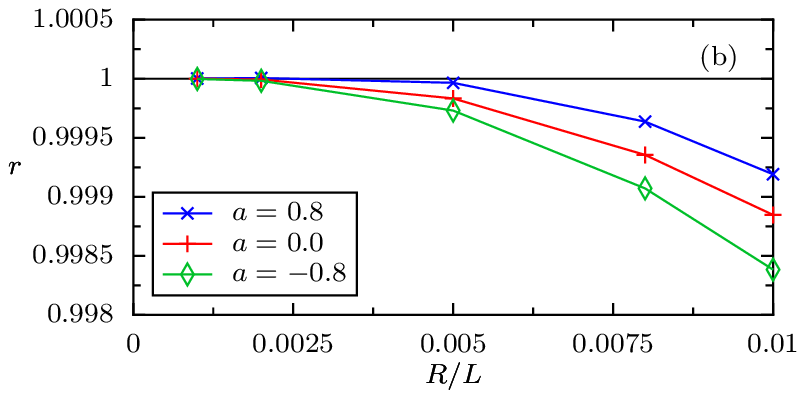}\medskip\par
  \includegraphics{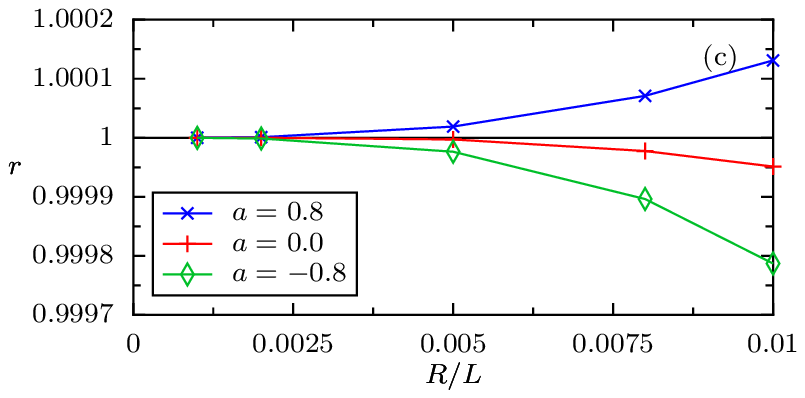}%
  \caption{(Color online) The relative accumulation~$r$ perpendicular to the
  main flow direction for flows with no-slip boundary conditions. Values smaller
  than unity correspond to an accumulation of particles below the separating
  streamline, near the straight wall. In panel~(a) the temperature is varied for
  a fixed particle radius $R=10^{-2}L$. In panels (b)~and (c) the particle
  radius is varied for two selected temperatures, $\Delta
  p/T=3.3\times10^{-7}$~and $3.3\times10^{-8}$, respectively.}%
  \label{fig:noslip_accum}
\end{figure}%

Figure~\ref{fig:noslip_accum} displays the resulting relative accumulation for
flows with two no-slip boundaries in three different channel geometries,
indicated by different values of the asymmetry parameter~$a$. The upper panel
shows the ratio~$r$ as a function of the inverse temperature. The first
observation is that the effect vanishes for vanishing driving -- or
equivalently, for infinite temperature. For the majority of the parameter values
the result is smaller than unity. This corresponds to an \emph{accumulation of
particles at the side of the straight wall}. This appears as a general tendency,
which was found also for other shapes. The relative accumulation effect is at
most one percent for the smallest temperature that was used in the calculation.
As expected, it vanishes for very small driving strengths~$\Delta p/T$. In order
to demonstrate that this small effect is not an artifact of the numerical
calculation, also the flows in the mirrored channels with inverted pressure
differences were considered. For symmetry reasons both configurations yield
identical accumulation ratios. The numerical differences between these are found
to be smaller than the line thickness in Fig.~\ref{fig:noslip_accum}. For two
temperature values, also the particle radius was varied. Panels (b)~and (c) of
Fig.~\ref{fig:noslip_accum} depict the ratio~$r$ as a function of the radius.
Again, the effect vanishes with vanishing radius. This is the expected behavior,
because a~point-particle can come arbitrarily close to the physical boundary.

An interesting aspect of the accumulation results in Fig.~\ref{fig:noslip_accum}
is the fact that the channel with $a=0.8$ behaves differently from the one
with~$a=-0.8$. We find as a general tendency that the channel which suddenly
widens and slowly narrows~($a=-0.8$) in flow direction yields better
accumulations than the suddenly narrowing one. The channel with the sinusoidal
shape~($a=0$) yields accumulation results somewhere in between.

\subsection{Accumulation inversion}

Another remarkable property of the accumulation mechanism can be observed for
the flow in the channel with asymmetry parameter~$a=0.8$. Here, we find values
of~$r$ also being larger than unity. Particles in this parameter regime are
transported towards the curved boundary rather than towards the straight one.
The occurrence of both values corresponds to an \emph{inversion of the transport
direction}. Panel~\ref{fig:noslip_accum}c confirms that the values larger than
unity, which we found in~\ref{fig:noslip_accum}a, persist also for several
smaller radii. Note that for each radius, a different numerical mesh was used in
order to represent the correct boundary~$\partial\Omega^\prime$. The inversion
found here is not readily usable for sorting particles because it takes place
upon varying the parameter $\Delta p/T$ and not the radius~$R$. Moreover, the
effect seems far too small to be of experimental relevance. Still, the occurrence
of accumulation inversion is an interesting effect.
\begin{figure}%
  \centering
  \includegraphics{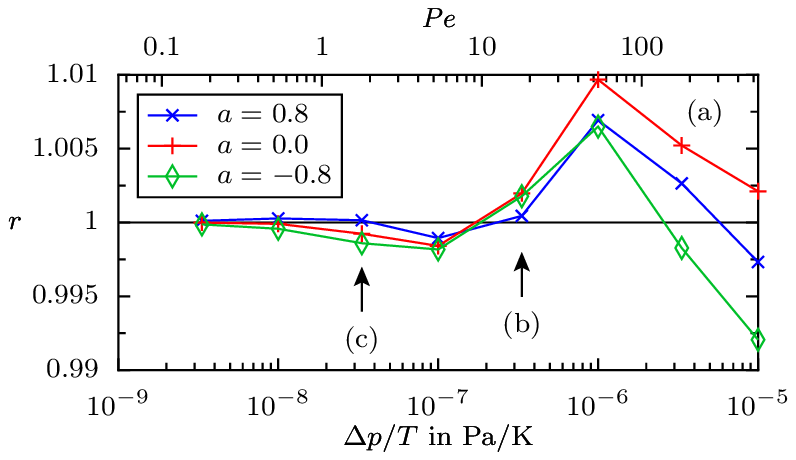}\medskip\par
  \includegraphics{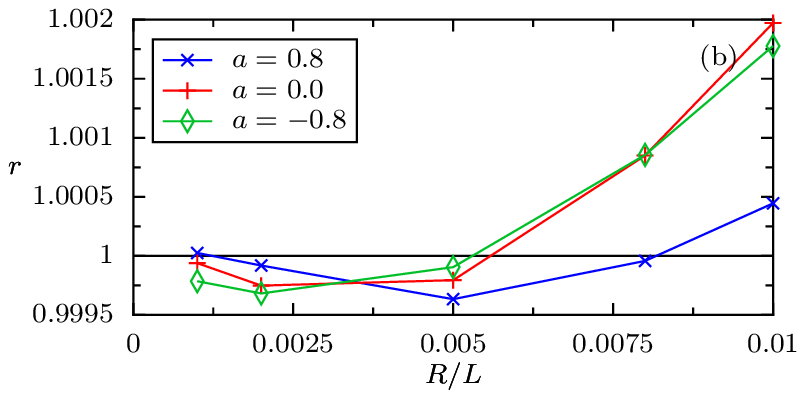}\medskip\par
  \includegraphics{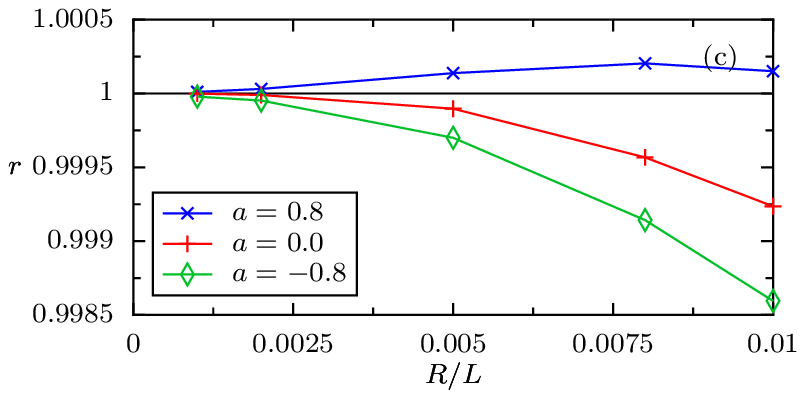}%
  \caption{(Color online) The same as Fig.~\ref{fig:noslip_accum} for a flow
  with perfect-slip boundary condition at the straight wall.}%
  \label{fig:botslip_accum}
\end{figure}%

Employing a perfect-slip boundary condition at the straight wall leads to a
qualitatively different parameter dependence of the accumulation.
Fig.~\ref{fig:botslip_accum} depicts the relative accumulation for the same
driving parameters as Fig.~\ref{fig:noslip_accum}. Again, the accumulation
vanishes for vanishing driving and for vanishing particle radius. The direction
of the accumulation depends now much stronger on the driving strength~$\Delta p/T$.
As a qualitative picture, we here observe a particle of a given size to be
pushed towards the curved or towards the straight wall, depending on the
pressure difference. This accumulation inversion can also be found in
Fig.~\ref{fig:botslip_accum}b as a function of the particle radius. Thus, we may
find particles with different radii accumulated in the two different parts of the
channel. For small particles this effect though is small.

\section{Conclusions}
We analyzed the accumulation of small spherical particles in two-dimensional
fluid flows confined to channels with one curved boundary. It turned out that
accumulation due to volume effects is impossible if the flow is driven by a
pressure difference or a homogeneous force. In this case, only boundary
accumulation effects are possible. For that to happen the normal projection of
the particle drift at the boundary must not vanish. Otherwise, only the trivial
uniform distribution of particles is achieved. In order to simplify the
numerical calculations we employed a hard-core interaction model for the
intricate interaction between a particle and a wall. Both imply non-vanishing
normal components of the drift velocity near the boundary, thus giving rise to a
small accumulation effect, which scales with the particle radius.

We compared no-slip and perfect-slip boundary conditions for the velocity fields
in the channels. The latter is a simplification of the free-surface boundary
condition which is realized in some experiments. We observed that the type of
the boundary condition on the flow has severe implications on the resulting
accumulation pattern of immersed particles. Free surfaces generally caused
larger accumulation effects showing also qualitatively different and interesting
properties, such as an inversion of the accumulation direction, depending on the
radius of the considered particle. This inversion might be put to beneficial use
for a separation of particles of varying size.

\appendix

\textbf{Acknowledgments}.
We thank Dr. Uwe Thiele for valuable comments on the manuscript.
Financial support is gratefully acknowledged from the \textit{Deutsche
Forschungsgemeinschaft} (DFG) via \textit{Sonderforschungsbereich~486, Projekt
B13}, the DFG grants HA\,1517/25-2 and HA\,1517/29-1, and also by the German
Excellence Initiative via the \textit {Nanosystems Initiative Munich (NIM)}.

\section{Velocity fields}
\label{appendix} The following three velocity fields are relevant in Fax\'en's
theorem to express the motion of a spherical particle in an external velocity
field: First, in absence of the particle, there is the unperturbed velocity
field solving the Stokes equations \eqref{eq:stokes} and~\eqref{eq:incompress}.
The flow field satisfies the integro-differential equation \cite{Pozrikidis92},
\begin{equation}
  \label{eq:integroeq}
  v_k(\bfy) = \oint\limits_{\partial\Omega} dA(\bfx)\,N_j(\bfx) \bigl[
        \calK_{ik}(\bfx-\bfy)\,\stress_{ij}(\bfx)
      - \calT_{ijk}(\bfx-\bfy)\,v_i(\bfx)
      \bigr],
\end{equation}
where the external conservative force is taken into account by defining a
modified pressure containing the potential of the force. The kernels $\calK$ and
$\calT$ may be chosen to be the Green functions for the unbounded Stokes
problem,
\begin{align}
  \calK_{ik}(\bfr)
  &= \frac{1}{8\pi\viscosity}\Bigl(\frac{\delta_{ik}}{r} +
  \frac{r_i r_k}{r^3}\Bigr),\\
  \calT_{ijk}(\bfr) &= -\frac{3}{4\pi} \frac{r_i r_j r_k}{r^5}.
\end{align}
The transformation of the Stokes equations into an equation on the boundary only
is essential for Fax\'en's theorem. This implies that only conservative forces
may drive the fluid.

The second flow field which is important is the true velocity field in presence
of the particle. We denote it with~$\bar{\bar\bfv}$. It satisfies a similar
relation as \eqref{eq:integroeq}, but the particle cuts a spherical region out
of the domain $\Omega$. The boundary integral is then carried out also over the
surface of the particle,
\begin{equation}
  \label{eq:integroeq_bar}
  \bar{\bar{v}}_k(\bfy) = \oint\limits_\scriptbox{1em}{\partial\Omega\cup S_R(\bfX)}
      dA(\bfx)\,N_j(\bfx) \bigl[
        \calK_{ik}(\bfx-\bfy)\,\bar{\bar{\stress}}_{ij}(\bfx)
      - \calT_{ijk}(\bfx-\bfy)\,\bar{\bar{v}}_i(\bfx)
      \bigr].
\end{equation}
The third relevant velocity field is the one that is used in Fax\'en's theorem.
It must be evaluated at the center of the sphere. This effective velocity
field~$\tilde\bfv$ is obtained by carrying out the integral
in~\eqref{eq:integroeq_bar} over the particle surface explicitly as
\begin{equation}
  \label{eq:vtilde}
  \tilde{v}_k(\bfy) := \oint\limits_{\partial\Omega} dA(\bfx)\,N_j(\bfx) \bigl[
        \calK_{ik}(\bfx-\bfy)\,\bar{\bar{\stress}}_{ij}(\bfx)
      - \calT_{ijk}(\bfx-\bfy)\,\bar{\bar{v}}_i(\bfx)
      \bigr].
\end{equation}
For the details of this calculation see Ref.~\cite{Schindler06}. This effective
velocity field is is then defined in the whole domain~$\Omega$ and can be
evaluated at the center of the particle. It is used for expressing the force on
a particle in Eq.~\eqref{eq:faxen}. Note that the boundary values of the true
velocity field $\bar{\bar\bfv}$ enter the definition of $\tilde\bfv$. As a
solution of the Stokes equation with appropriate pressure field, it is
solenoidal. We make use of this property in Sec.~\ref{sec:volume}.


\end{document}